\documentclass[fleqn,12pt,twoside]{article}
\usepackage{espcrc1,psfig}
\usepackage{graphicx}
\usepackage[figuresright]{rotating}

\newcommand{\gton}{\mathrel{\lower.5ex \hbox{$\stackrel{> }
 {\scriptstyle \sim}$}}}
\newcommand{\lton}{\mathrel{\lower.5ex \hbox{$\stackrel{< }
 {\scriptstyle \sim}$}}}

\newcommand{\AmS}{{\protect\the\textfont2
  A\kern-.1667em\lower.5ex\hbox{M}\kern-.125emS}}

\title{High-$p_T$ Pion Quenching versus anti+Baryon Enhancement 
       in Nucleus-Nucleus Collisions}

\author{Ivan Vitev\address[CU]{Department of Physics, 
        Columbia University, \\ 
        538 West 120-th Street, New York, NY 10027, USA} and 
        Miklos Gyulassy\addressmark[CU]\thanks{This work is 
supported by the Director, Office of Science, 
Office of High Energy and Nuclear Physics,
Division of Nuclear Physics, of the U.S. Department of Energy
under Grant No. DE-FG02-93ER40764.} }
       
\begin{document}

\maketitle

\begin{abstract}
Jet quenching of quarks and gluons in central $A+A$ collisions 
suppresses the high-$p_T$ meson production and exposes novel 
baryon dynamics that we attribute to (gluonic) baryon junctions. 
This mechanism predicts baryon enhancement in a finite moderate-$p_T$ 
window that decreases with increasing impact parameter. 
We also extend recent calculations of the transverse momentum 
behavior of $p/\pi^+$ and $\bar{p}/\pi^-$ to  other baryon species 
and show that a similar pattern is expected for the $K^-/\Lambda$ 
and $K^+/\bar{\Lambda}$ ratios. Within the framework of the 
model constant $\bar{p}/p$ and $\bar{\Lambda}/\Lambda$ 
ratios are found in central $Au+Au$ collisions at RHIC energies  up to 
$p_T \simeq 4 - 5$~GeV.  
\end{abstract}

\section{INTRODUCTION}

One of the unexpected results reported by PHENIX and STAR during the
first year RHIC run at $\sqrt{s}_{NN}=130$~GeV was that in contrast 
to the strong $\pi^0$ quenching for $2\; {\rm GeV} < p_T < 5\;{\rm GeV}$,  
the corresponding charged hadrons  were found to be  
suppressed by only a  factor $\sim 2-2.5$. Even more surprisingly, 
the identified particle spectra analysis at PHENIX  
suggests that $R_B(p_T)=\bar{p}/\pi^-, \, p/\pi^+  \gton 1$ for
$p_T > 2$~GeV.  Thus, baryon and antibaryon production may in 
fact  dominate the  moderate- to high-$p_T$  hadron flavor 
yields~\cite{baryons130}.

These and other data point to novel baryon transport 
dynamics  playing role in nucleus-nucleus reactions.
An important indicator of this is the high valence proton 
rapidity density at midrapidity ($y=0$). More recent results 
corroborate the non-perturbative baryon production hypothesis 
through equally abundant  $\Lambda$ and $\bar{\Lambda}$ 
production~\cite{baryons130}. It has also been observed that the
mean transverse momentum $\langle p_T \rangle_B$ for various baryon
and antibaryon species is approximately constant and deviates from the 
common hydrodynamic flow systematics of soft hadron production in 
$A+A$ collisions. This provides strong motivation to explore 
alternative physical mechanisms that may give insight into the 
anomalous anti+baryon behavior at RHIC. 

Identified particle analyses from the recent $\sqrt{s}_{NN}=200$~GeV RHIC 
run find similar puzzling features of moderate-$p_T$ baryon spectra that
have to be understood in the context of even stronger   
quenching of neutral pions~\cite{baryon&supp200}.

\begin{figure}[htb]
\begin{minipage}[t]{77.5mm}
\psfig{figure=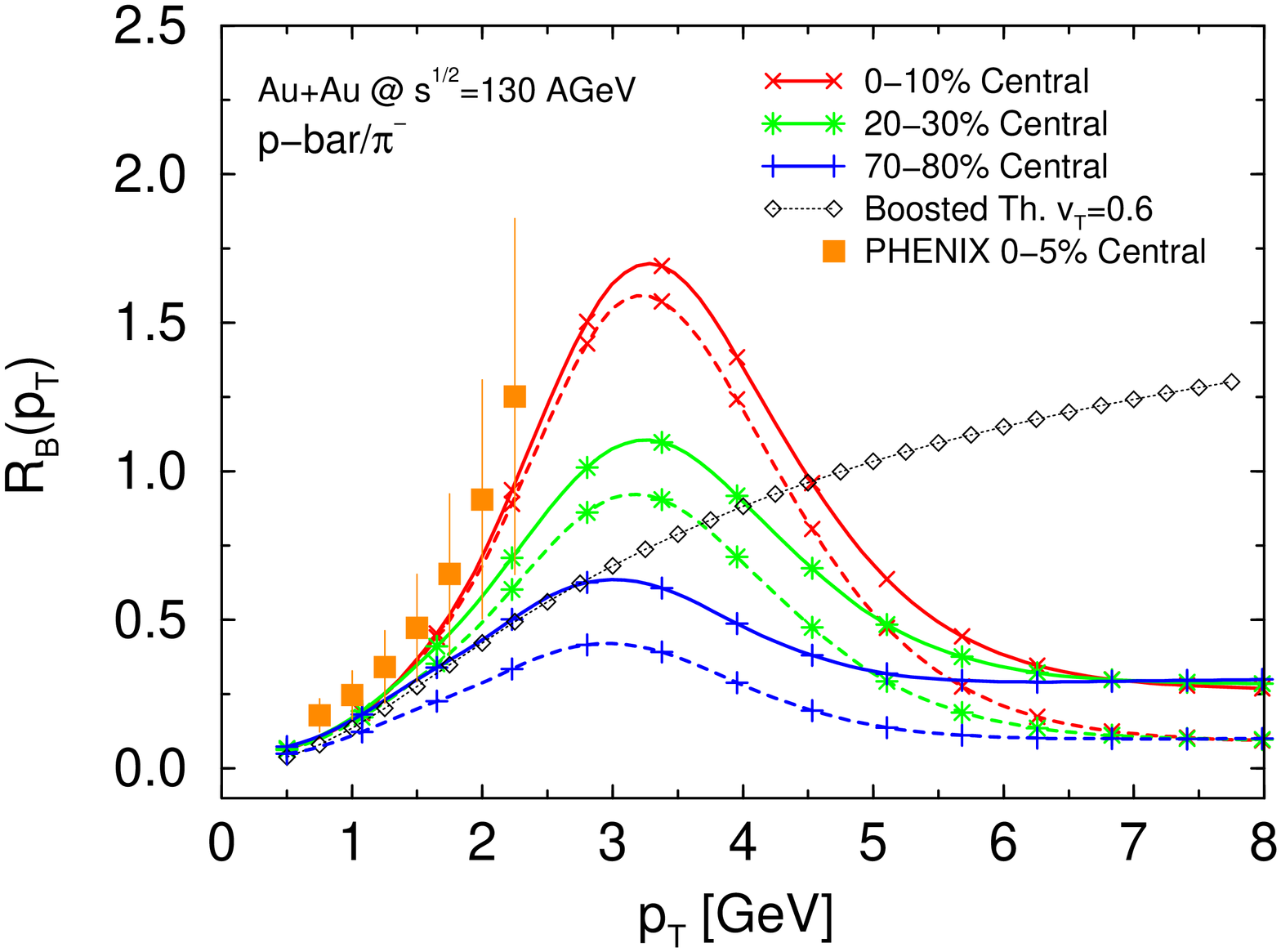,height=2.4in,width=3.in,angle=-90} 
\vspace*{-0.9cm}
\caption{The $\bar{p}/\pi^-$ ratio versus $p_T$ for 3 centrality classes 
at $\sqrt{s}_{NN}=130$~GeV. $N_{part}$(solid) versus $N_{bin}$(dashed)
geometry is shown. Boosted thermal source computation and 
ratio of {\em fits} to PHENIX data are shown  for comparison.}
\label{fig:ptopi}
\end{minipage}
\hspace{\fill}
\begin{minipage}[t]{77.5mm}
\psfig{figure=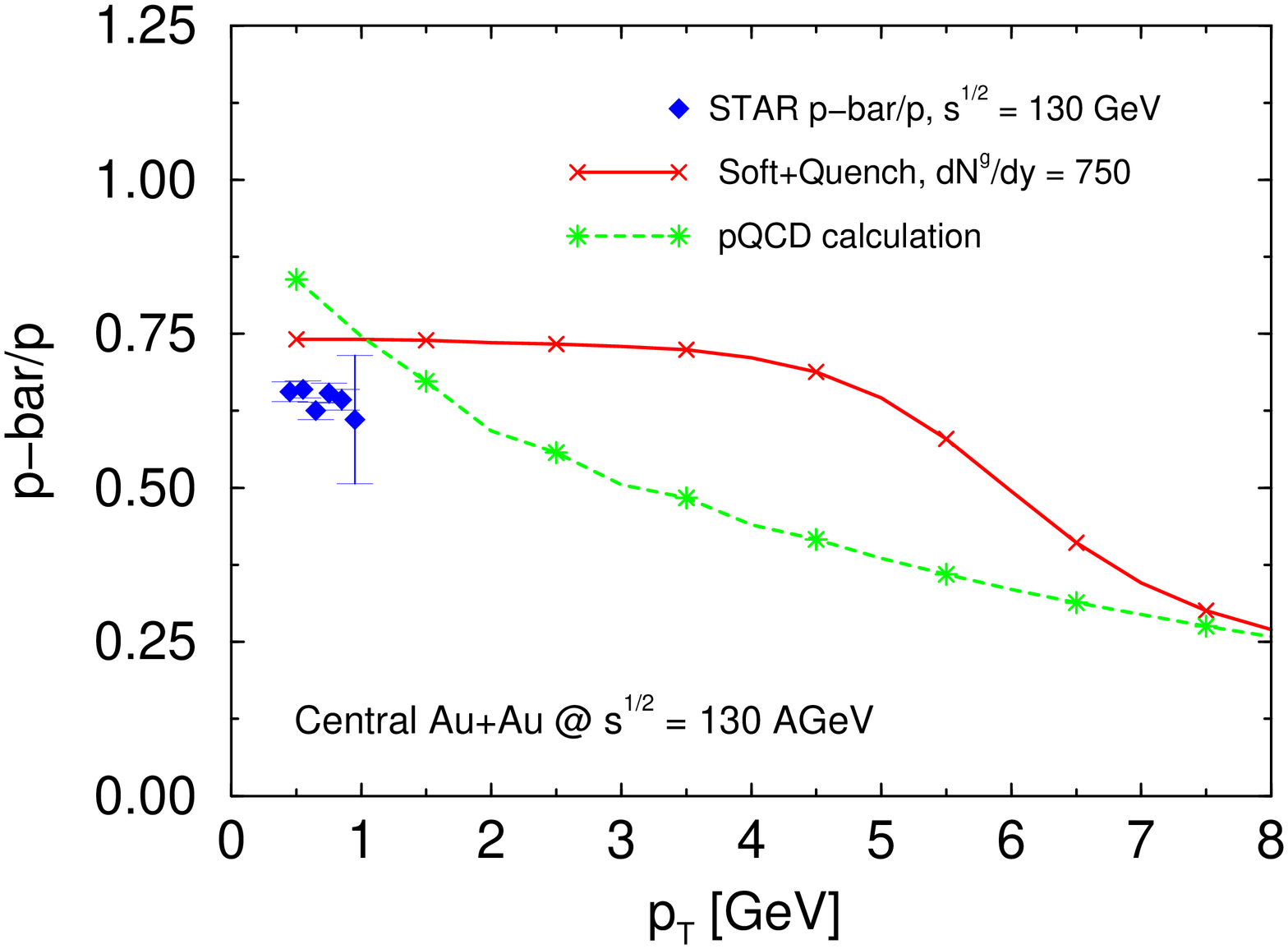,height=2.4in,width=3.in,angle=-90} 
\vspace*{-0.9cm}
\caption{The $\bar{p}/p$ ratio versus $p_T$ for central $Au+Au$ at 
$\sqrt{s}_{NN}=130$~GeV. Comparison between a PQCD calculation and 
a quenched+baryon junction scenario. Central 10\% data from STAR.}
\label{fig:toosmall}
\end{minipage}
\end{figure}

\section{BARYON PRODUCTION AND TRANSPORT MECHANISM}

A topological non-perturbative baryon production and transport mechanism was 
originally proposed by Veneziano and Rossi in elementary nucleon-nucleon
collisions~\cite{junction}. It has recently been successfully generalized 
and implemented for the case of  nucleus-nucleus collisions~\cite{junction}. 
Its phenomenological applications are currently based on Regge theory, where
a Regge trajectory $J=\alpha(0)+\alpha^\prime(0) M^2$ is specified by its
intercept $\alpha(0)$ and slope $\alpha^\prime(0)$. It has been 
argued~\cite{junction} that $\alpha_J(0) \simeq 0.5$ and  $\alpha^\prime_J(0) 
\simeq 1/3 \,  \alpha^\prime_R(0)$. Regge theory gives exponential 
rapidity correlations, which in the presence of two sources  
(at $\pm Y_{\max}$) lead to net baryon rapidity density in central 
$A+A$ collisions of the form: 
\begin{equation} 
\frac{dN^{B-\bar{B}}}{dy} = (Z+N)(1-\alpha_J(0)) 
\frac{\cosh (1-\alpha_J(0)) y} {\sinh (1-\alpha_J(0))Y_{\max}} \, .
\label{btrans}
\end{equation}
It is evident from Eq.(\ref{btrans}) that the net baryon distribution 
integrates to $2A$ and in going to peripheral reactions scales as $N_{part}$.
At RHIC energies of $\sqrt{s}=130(200)$~GeV,  
corresponding to $Y_{\max} = 4.8(5.4)$,  in central reactions  
$dN^{B-\bar{B}}/{dy} = d(p-\bar{p})/dy+ d(n-\bar{n})/dy + 
d(\Lambda -\bar{\Lambda})/dy + \cdots \simeq 18(13.5) $. The relative 
contribution of each baryon species can be evaluated from isospin 
symmetry and strangeness conservation (via comparison to midrapidity kaon 
production). 

Hadronic transport in small-to-moderate $p_T$ is effectively controlled 
by the slope of the Regge trajectory. This would suggest that the 
baryon/meson mean inverse slopes in a phenomenological $p_T$-exponential 
($\sim e^{-p_T/T}$) soft particle production model are related as 
$T_B:T_M \simeq \sqrt{3}:\sqrt{2}$. Soft pion production, however, is  
largely  dominated by resonance decays, where the cumulative effect 
from the random walk in $p_T$ due to string braking is destroyed. This 
leads to the relation 
$\langle p_T \rangle_\pi : \langle p_T \rangle_K : \langle p_T \rangle_B
\simeq 1: (1 \div \sqrt{2}): \sqrt{3} $ (220~MeV : 275 MeV : 
380 MeV)~\cite{gvbaryon}. One also notes 
that in the limit of pair production dominated by junction-antijunction 
loops (which we consider here)  the transverse momentum distribution of
antibaryons closely resembles that of baryons 
(with $\langle p_T \rangle_{\bar{B}} = \langle p_T \rangle_B $).

The high-$p_T$ part of the hadron spectra is computed from leading 
order PQCD augmented by the effects of initial multiple scattering 
(Cronin), nuclear shadowing, and  parton energy loss. The most important 
modification to the perturbative scheme comes from the final state 
gluon bremsstrahlung  computed here via the Gyulassy-Levai-Vitev (GLV) 
formalism~\cite{eloss}. Its centrality dependence determines to a large
extent (especially for pions) the centrality dependence of the 
baryon/meson ratios. Similar technique  has  been used to estimate 
the  $A$-induced initial  parton  broadening~\cite{eloss}.

\begin{figure}[t]
\begin{minipage}[t]{77.5mm}
\psfig{figure=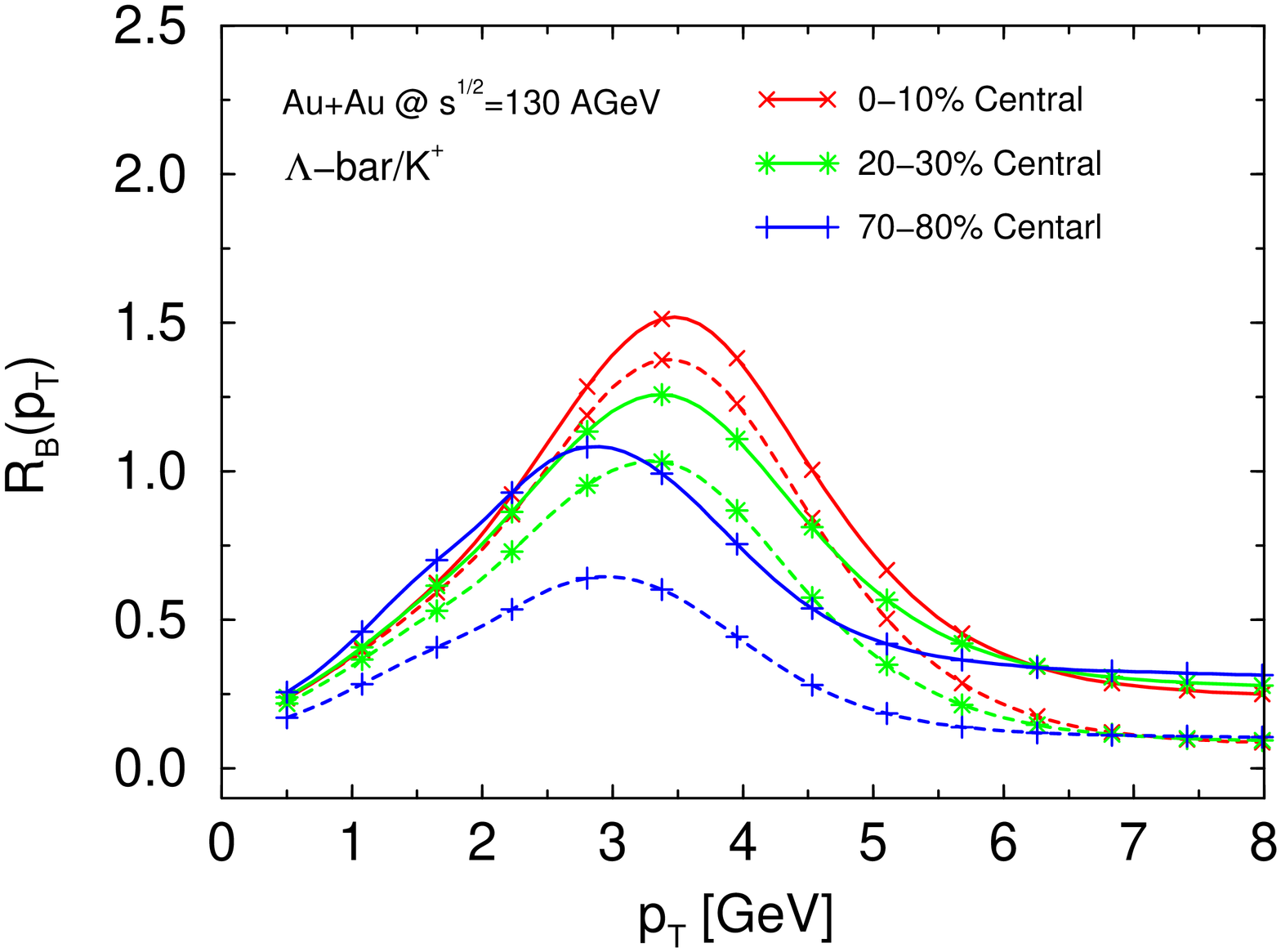,height=2.4in,width=3.in,angle=-90} 
\vspace*{-0.9cm}
\caption{Same centrality classes as in Fig.~1 for 
$\bar{\Lambda}/K^+$ versus $p_T$. $Au+Au$  reactions 
at $\sqrt{s}_{NN}=130$ GeV. }
\label{fig:ptopi}
\end{minipage}
\hspace{\fill}
\begin{minipage}[t]{77.5mm}
\psfig{figure=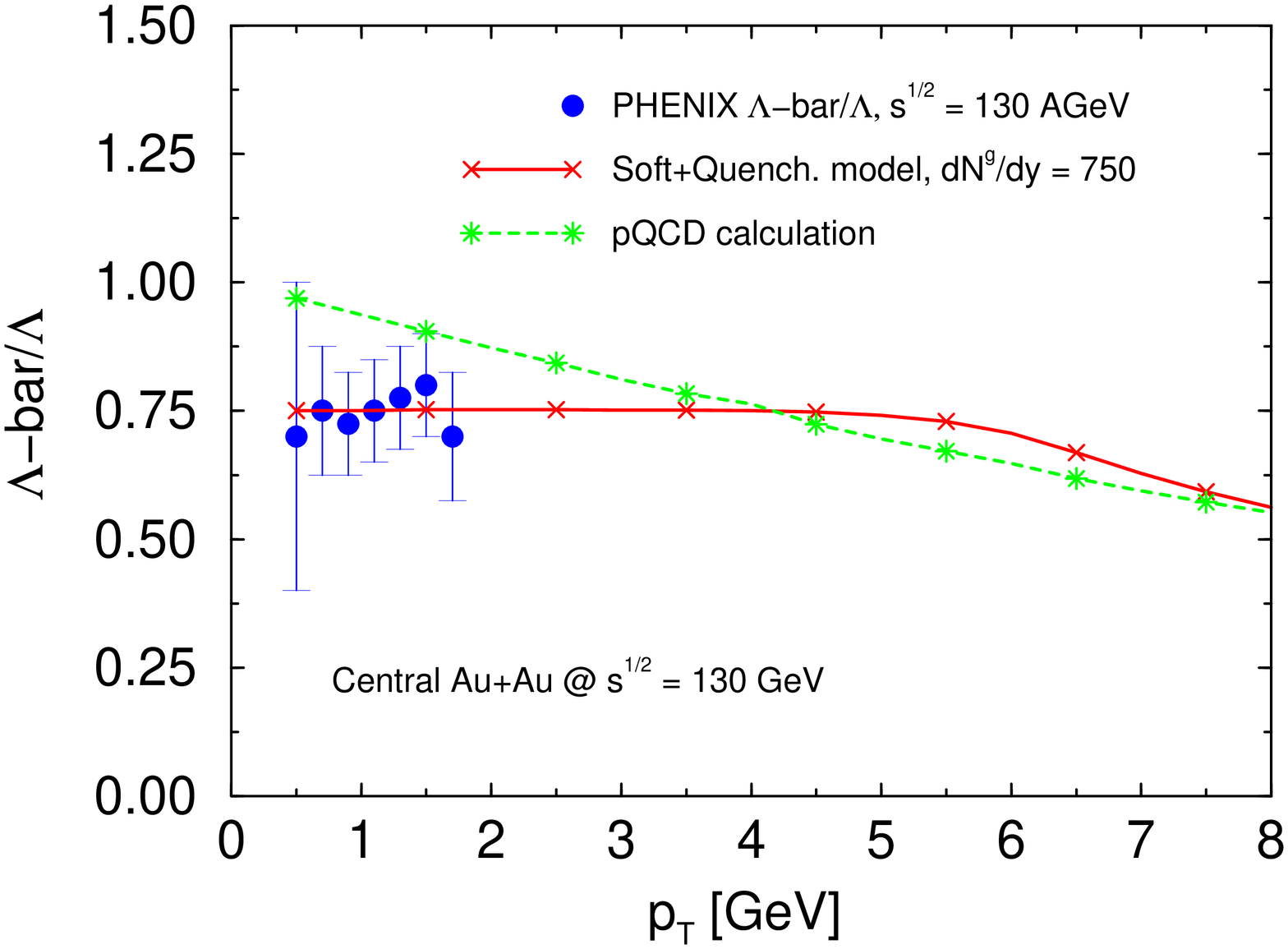,height=2.4in,width=3.in,angle=-90} 
\vspace*{-0.9cm}
\caption{Same as Fig.~2 for $\bar{\Lambda}/\Lambda$ in central $Au+Au$ at 
$\sqrt{s}_{NN}=130$ GeV. Minimum-bias data from PHENIX.}
\label{fig:toosmall}
\end{minipage}
\end{figure}

\begin{figure}[t]
\begin{minipage}[t]{77.5mm}
\psfig{figure=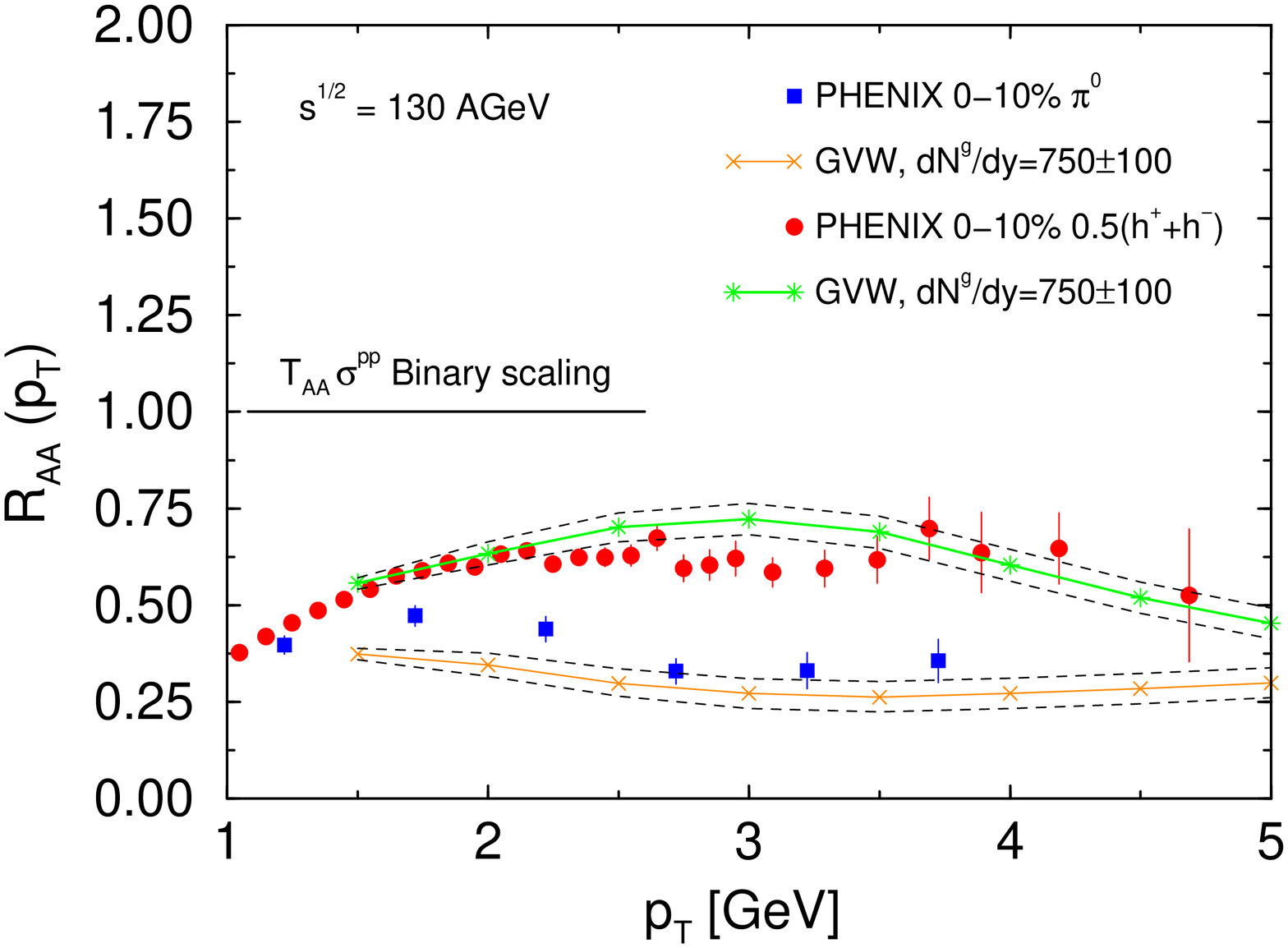,height=2.4in,width=3.in,angle=-90} 
\vspace*{-0.9cm}
\caption{Suppression  of neutral pions and inclusive 
charged hadrons relative to the binary collision scaled $p+p$ result.
Central $Au+Au$ at  $\sqrt{s}_{NN}=130$ GeV.}
\label{fig:ptopi}
\end{minipage}
\hspace{\fill}
\begin{minipage}[t]{77.5mm}
\psfig{figure=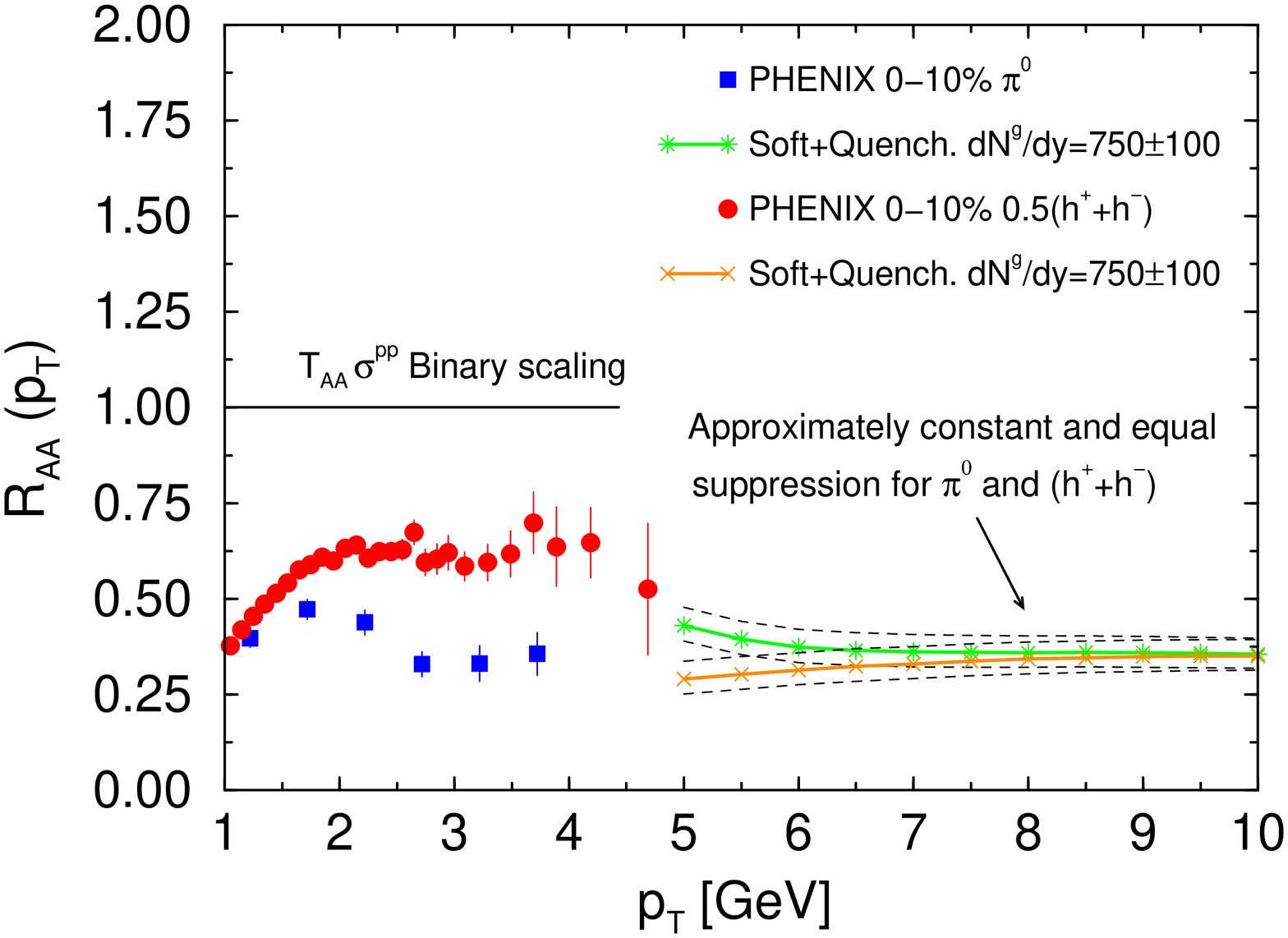,height=2.4in,width=3.in,angle=-90} 
\vspace*{-0.9cm}
\caption{Same as Fig.~5 extended to large transverse momenta.   
Note $R_{AA}( p_T \gton 5-6\; {\rm GeV} ) \approx const.$ }
\label{fig:toosmall}
\end{minipage}
\end{figure}

\section{RESULTS AND CONCLUSIONS}

We have studied the interplay between a novel baryon production and transport
mechanism~\cite{junction} (in rapidity $y$  and moderate $p_T$) and 
the  suppression of  particle  spectra as a result of the medium 
induced non-Abelian energy loss of jets~\cite{eloss} 
in $Au+Au$ reactions at $\sqrt{s}_{NN}=130$~GeV.  We find an enhanced 
baryon/meson ratio $R_B(p_T) \gton 1$ that decreases with centrality 
in a finite $p_T$ window as illustrated in Figs.~1 and 3. For 
$p_T > 5$~GeV the ratios reduce below unity and follow the perturbative
calculation.  We have extended recent  studies of the $p_T$-differential 
$p/\pi^+$, $\bar{p}/\pi^-$ ratios~\cite{gvbaryon}  (Fig.~1) to    
$K^-/\Lambda$, $K^+/\bar{\Lambda}$ (Fig.~3). Similar enhancement 
pattern if found at moderate  $2\; {\rm GeV} \leq p_T 
\leq 5$~GeV. Its centrality dependence is predicted to be weak.  
Figs.~2 and~4 show that in central reactions the 
predicted $\bar{p}/p$ and $\bar{\Lambda}/\Lambda$  ratios 
are $p_T$-independent up to $p_T \simeq 4-5$~GeV. 
At larger transverse momenta they decrease and approach 
perturbative estimates~\cite{pbartop}. We propose that the 
non-perturbative baryon dynamics may extend to $p_T \simeq 5$~GeV.

The enhanced baryon and antibaryon production at moderate transverse
momenta accounts for the difference in the  suppression 
factor $R_{AA}(p_T)$ of inclusive charged hadrons and neutral 
pions~\cite{gvbaryon} (see Fig.~5).
We have demonstrated that when the non-perturbative baryon contribution   
becomes small $R_{AA}(p_T)$ is approximately equal for all hadron species.
This is illustrated  in Fig.~6 for  $p_T \geq 5$~GeV. For the first RHIC 
run at $\sqrt{s}_{NN}=130$~GeV up to $p_T \simeq 10$~GeV the suppression 
factor is found to have a value  $R_{AA}(p_T) \simeq 0.3 $ and is almost 
independent of the transverse momentum~\cite{gvbaryon}.  

 
We have performed PQCD calculations at the highest RHIC energy of
 $\sqrt{s}_{NN}=200$~GeV that 
also predict a constant suppression factor $R_{AA}(p_T)$.
The detailed $R_{AA}(p_T)$  behavior is shown to be a consequence 
of the combination of Cronin effect, shadowing, energy loss, and 
the computable shape of the underlying jet distributions.


\begin{thebibliography}{9}


\bibitem{baryons130} J.~Velkovska,
Nucl. Phys. A {\bf 698}, 507 (2002); K. Adcox {\em et al.}, 
Phys. Rev. Lett. {\bf 88}, 242301 (2002); nucl-ex/0204007;
N.~Xu and M.~Kaneta, Nucl. Phys. A {\bf 698}, 306 (2002);
C.~Adler {\em et al.},  Phys. Rev. Lett. {\bf 87}, 262302 (2001); 
nucl-ex/0203016. 


\bibitem{baryon&supp200}
S. Mioduszewski, these proceedings; G. Kunde, these proceedings; 
D. d'Enterria, these proceedings; J.~Jiangyong, these proceedings.


\bibitem{junction} G.C.~Rossi and G.~Veneziano, Nucl. Phys. B {\bf 123}, 
507 (1977);  Phys. Rept. {\bf 63}, 153 (1980); 
D.~Kharzeev, Phys. Lett. B {\bf 378}, 238 (1996);
S.E.~Vance, M.~Gyulassy, X.-N.~Wang,
 Phys. Lett. B {\bf 443}, 45 (1998); S.E.~Vance and M.~Gyulassy,
 Phys. Rev. Lett. {\bf 83}, 1735 (1999).


\bibitem{gvbaryon}I.~Vitev and M. Gyulassy, 
Phys. Rev. C {\bf 65}, 041902 (2002); hep-ph/0108045;
I.~Vitev, M. Gyulassy and P.~Levai, hep-ph/0109198.


\bibitem{eloss} 
M.~Gyulassy, P.~Levai, and  I.~Vitev, Nucl. Phys. B {\bf 571},  
197 (2000); Phys.\ Rev.\ Lett.\  {\bf 85}, 5535 (2000);
Nucl.\ Phys.\ B {\bf 594}, 371 (2001); 
Phys. Lett. B {\bf 538}, 282 (2002); Phys. Rev. D {\bf 66} 
014005 (2002).

\bibitem{pbartop}
X.-N.~Wang, Phys. Rev. C {\bf 58}, 2321 (1998); X.-F. Zhang, G. Fai, 
and P. Levai, hep-ph/0205008. 


\end{thebibliography}
\end{document}